\documentclass[twocolumn,english,aps,prl,reprint,superscriptaddress]{revtex4-2}
\usepackage[T1]{fontenc}
\usepackage[latin9]{inputenc}
\usepackage{amsmath}
\usepackage{amssymb}
\usepackage{graphicx}
\usepackage{wasysym}
\usepackage{color}

\makeatletter
\@ifundefined{textcolor}{}
{%
 \definecolor{BLACK}{gray}{0}
 \definecolor{WHITE}{gray}{1}
 \definecolor{RED}{rgb}{1,0,0}
 \definecolor{GREEN}{rgb}{0,1,0}
 \definecolor{BLUE}{rgb}{0,0,1}
 \definecolor{CYAN}{cmyk}{1,0,0,0}
 \definecolor{MAGENTA}{cmyk}{0,1,0,0}
 \definecolor{YELLOW}{cmyk}{0,0,1,0}
 \definecolor{figa}{rgb}{0.7143, 0.7143, 0.7143}
 \definecolor{figb}{rgb}{0.4416, 0.7490, 0.4322}
 \definecolor{figc}{rgb}{0.3639, 0.5755, 0.7484}
 \definecolor{figd}{rgb}{0.9153, 0.2816, 0.2878}
 \definecolor{fige}{rgb}{1.0000, 0.5984, 0.2000}
 \definecolor{figf}{rgb}{0.6365, 0.3753, 0.6753}
}

\usepackage{pslatex}

\usepackage[space]{grffile}

\graphicspath{{Figures/}
	{Figures/Figure 1}
	{Figures/Figure 2}
	{Figures/Figure 3}
	{Figures/Figure 4}
}

\usepackage{babel}

\makeatother

\begin{document}

\title{Magneto-Optical Trapping and Sub-Doppler Cooling of a Polyatomic Molecule}

\author{Nathaniel B. Vilas}
\email{vilas@g.harvard.edu}
\affiliation{Department of Physics, Harvard University, Cambridge, MA 02138, USA}
\affiliation{Harvard-MIT Center for Ultracold Atoms, Cambridge, MA 02138, USA}

\author{Christian Hallas}
\affiliation{Department of Physics, Harvard University, Cambridge, MA 02138, USA}
\affiliation{Harvard-MIT Center for Ultracold Atoms, Cambridge, MA 02138, USA}

\author{Lo\"ic Anderegg}
\affiliation{Department of Physics, Harvard University, Cambridge, MA 02138, USA}
\affiliation{Harvard-MIT Center for Ultracold Atoms, Cambridge, MA 02138, USA}

\author{Paige Robichaud}
\affiliation{Department of Physics, Harvard University, Cambridge, MA 02138, USA}
\affiliation{Harvard-MIT Center for Ultracold Atoms, Cambridge, MA 02138, USA}

\author{Andrew Winnicki}
\affiliation{Department of Physics, Harvard University, Cambridge, MA 02138, USA}
\affiliation{Harvard-MIT Center for Ultracold Atoms, Cambridge, MA 02138, USA}

\author{Debayan Mitra}
\altaffiliation{Current address: Department of Physics, Columbia University, New York, NY 10027, USA.}
\affiliation{Department of Physics, Harvard University, Cambridge, MA 02138, USA}
\affiliation{Harvard-MIT Center for Ultracold Atoms, Cambridge, MA 02138, USA}

\author{John M. Doyle}
\affiliation{Department of Physics, Harvard University, Cambridge, MA 02138, USA}
\affiliation{Harvard-MIT Center for Ultracold Atoms, Cambridge, MA 02138, USA}

\date{\today}

\begin{abstract}

We report magneto-optical trapping (MOT) of a polyatomic molecule, calcium monohydroxide (CaOH). The MOT contains $2.0(5)\times 10^4$ CaOH molecules at a peak density of $3.0(8)\times10^{6}$ cm$^{-3}$. CaOH molecules are further sub-Doppler laser cooled in an optical molasses, to a temperature of 110(4) $\mu$K. The temperatures and densities achieved here make CaOH a viable candidate for a wide variety of quantum science applications, including the creation of optical tweezer arrays of CaOH molecules. This work also suggests that laser cooling and magneto-optical trapping of many other polyatomic species \cite{isaev2016polyatomic, kozyryev2016proposal, dickerson2021franck} will be both feasible and practical.

\end{abstract}
\maketitle

Laser cooling and trapping are cornerstone experimental techniques of atomic and molecular physics \cite{phillips1998nobel, chu1998nobel}. The magneto-optical trap (MOT) is a central tool that enables rapid spatial compression and cooling of atoms and molecules to submillikelvin temperatures. MOTs of atoms have enabled groundbreaking advances, including Bose-Einstein condensation (BEC) \cite{cornell2002nobel, anderson1995observation, davis1995bose}, simulation of quantum many-body systems \cite{bloch2008many}, quantum computation with neutral atoms \cite{bernien2017probing, levine2019parallel}, and high-precision optical clocks \cite{ludlow2015optical}. More recently, MOTs of diatomic molecules \cite{barry2014magneto, truppe2017molecules, anderegg2017radio, collopy20183d} have opened access to new research with ultracold molecules, including quantum simulation \cite{micheli2006toolbox}, quantum computation \cite{demille2002quantum}, and searches for physics beyond the standard model (BSM) \cite{safronova2018search}.  The rapid cooling and compression provided by MOTs is highly desirable and has led to loading of molecules into optical tweezer arrays \cite{anderegg2019optical,zhang2021optical}.

Polyatomic molecules, as compared to diatomic molecules, have additional rotational and vibrational degrees of freedom that promise new possibilities in quantum simulation \cite{wall2013simulating, wall2015realizing, wall2014quantum}, quantum computation \cite{yu2019scalable}, ultracold collisions \cite{augustovicova2019collisions}, and BSM searches \cite{kozyryev2017precision, kozyryev2021enhanced}. Due to these distinct degrees of freedom, polyatomic molecules are also more challenging to cool and trap. Despite the challenges arising from molecular complexity, several techniques have previously demonstrated trapping and/or cooling of polyatomic molecules. CH$_3$ has been trapped in a permanent magnetic trap \cite{liu2017magnetic}, while trapped H$_2$CO molecules have been cooled below 1 mK by electro-optic Sisyphus cooling \cite{zeppenfeld2012sisyphus, prehn2016optoelectrical}. While polyatomic molecules including SrOH \cite{kozyryev2017sisyphus}, CaOH \cite{baum20201d}, YbOH \cite{augenbraun2019laser}, and CaOCH$_3$ \cite{mitra2020direct} have all been laser cooled in one dimension, polyatomic molecules have not previously been trapped and cooled in a MOT.

In this work, we demonstrate magneto-optical trapping and sub-Doppler laser cooling of calcium monohydroxide (CaOH) molecules. This is accomplished by optical photon cycling of CaOH sufficient to enable laser slowing, trapping in a MOT, and cooling to temperatures near the Doppler limit. To further cool, we apply a blue-detuned optical molasses, reaching temperatures below the Doppler limit. This work demonstrates trapping and cooling of a polyatomic molecule, as well as the photon cycling needed to perform high fidelity quantum state readout. The realized density and temperature represents a viable starting point for further laser cooling to lower temperatures \cite{cheuk2018lambda, caldwell2019deep} and the loading of CaOH molecules into optical dipole trap arrays \cite{anderegg2018laser, Langin2021, anderegg2019optical}.

Our experiment begins with a cryogenic buffer gas beam (CBGB) \cite{hutzler2012buffer} of CaOH molecules produced in a two-stage buffer gas cell cooled to $\sim$2 K. In brief, a metal calcium (Ca) target in the cell is ablated, producing Ca atoms that react in the gas phase with water (H$_2$O), which is introduced through a heated capillary into the cell. CaOH production via a Ca-H$_2$O reaction is enhanced by optically exciting the reactant Ca atoms to the metastable $4s4p\,^3P_1$ state \cite{jadbabaie2020enhanced, Supplemental}. CaOH molecules exit the cell with a peak forward velocity of $v_f \sim140$ m/s and a flux of $\sim$$5\times 10^{10}$ molecules/sr/pulse in the $N''=1$ rotational state used for laser cooling.

\begin{figure*}[ht]
    \centering
    \includegraphics[width = 2\columnwidth]{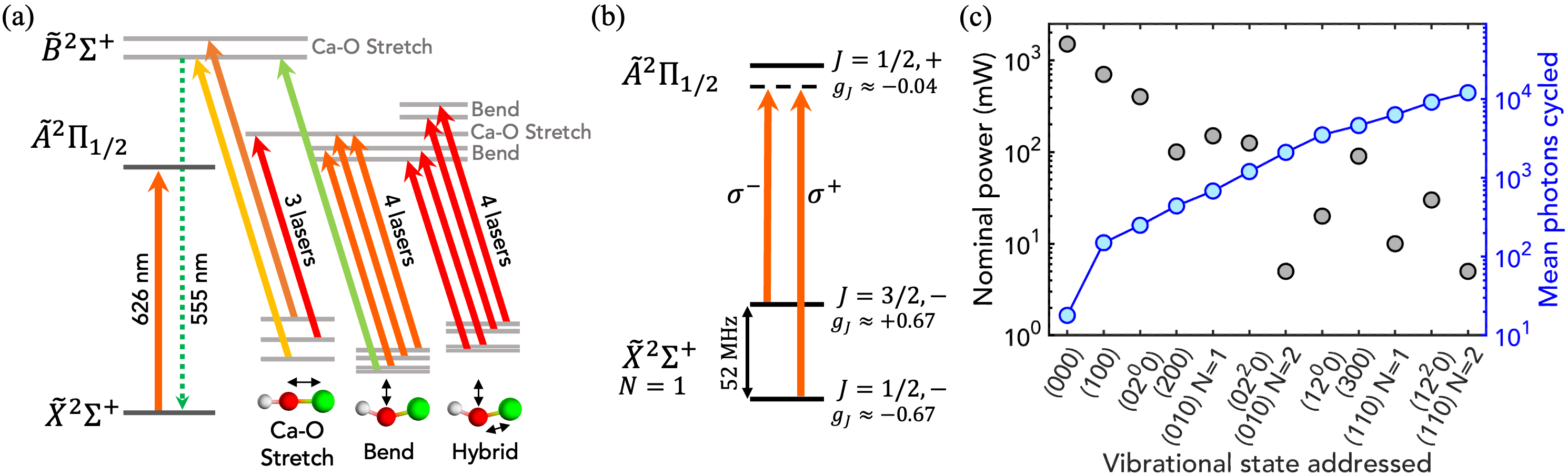}
    \caption{(a) Photon cycling scheme. The primary cycling transition for slowing and the MOT is the $\widetilde{X}\,^2\Sigma^+ \leftrightarrow \widetilde{A}\,^2\Pi_{1/2}$ transition at 626~nm. Spontaneous decay from the laser repumping the first Ca-O stretching mode is at 555~nm and is used for background free imaging. Decays to Ca-O stretch, bend, and hybrid vibrational modes in the electronic ground state are repumped though a number of excited vibronic states to attain a closure of $\sim$12,000 photons \cite{Supplemental}. (b) CaOH MOT polarization configuration. The MOT beams addressing the $J''=3/2$ and $J''=1/2$ ground states have opposite polarization due to their opposite $g$-factors. (c) Cycling laser properties. Left axis: Power of each slowing and repumping laser required to saturate the laser slowing and MOT efficacy. Right axis: Mean number of photons which can be scattered, by adding all of the preceding repumping lasers (from left-to-right), before $1/e$ loss of molecular population occurs \cite{baum2021establishing,zhang2021accurate}. Lasers are labeled by the vibrational state repumped, notated as $(v_1 v_2^\ell v_3)$, where $v_1$, $v_2$, and $v_3$ are the vibrational quantum numbers for the Ca--O stretch, Ca--O--H bend, and O--H stretch modes, respectively, and $\ell$ is the projection of the vibrational angular momentum onto the molecular axis.}
    \label{fig:Fig1}
\end{figure*}

As the CaOH molecules travel from the cell to the MOT region, which is 81 cm downstream, they are illuminated with counter-propagating white-light slowing lasers \cite{yeo2015rotational, hemmerling2016laser, truppe2017intense}, including the ``main'' 626 nm $\widetilde{A}\,^2\Pi_{1/2}(J'=1/2,p=+) \leftarrow \widetilde{X}\,^2\Sigma^+ (N''=1,p=-)$ cycling transition \cite{baum20201d}. Given the recoil velocity of $\hbar k/m \approx 11\times10^{-3}$ m/s (where $k=2\pi/(626 \text{ nm})$ and $m=57$ u is the mass of CaOH), $\sim$$10,000$ photons must be scattered per molecule in order to slow the velocity peak of the molecular beam to below the MOT capture velocity ($\lesssim$10 m/s) \cite{baum20201d}. Additional ``repumping'' lasers copropagate with the main cycling light and repump population that decays to three Ca--O stretch, three Ca--O--H bend, and three mixed character vibrational modes (Fig.\ref{fig:Fig1}(a)) \cite{baum2021establishing, zhang2021accurate}. Two more lasers are added for rotational repumping of vibrational modes with parity doublet structure \cite{baum2021establishing, Supplemental}. The suite of repumpers allows for $\sim$12,000 photons to be scattered before $1/e$ of the initial molecular population is lost to unaddressed vibrational states \cite{zhang2021accurate,Supplemental}. As we will describe, this number is amply sufficient for laser slowing and subsequent cooling and trapping in a MOT.

The main cycling and repumping lasers used for slowing are frequency-broadened by $\sim$300 MHz so that molecules with velocities as high as 140 m/s can be slowed to rest in the MOT region. The counter-propagating slowing light interacts with the CaOH molecular beam along the full slowing length and is focused from a 11.5 mm 1/$e^2$ Gaussian diameter at the MOT region to a 5.5 mm diameter at the buffer gas cell. The polarization of each slowing laser is rapidly switched at $\sim$1.4 MHz between two orthogonal linear polarizations using a Pockels cell, ensuring remixing of magnetic dark states in the cycling scheme over the full slowing length \cite{berkeland2002destabilization}. The minimum laser powers required to saturate the slowing effect on the molecules, as well as to optimize the trapping and cooling efficacy, are plotted in Fig. \ref{fig:Fig1}(c) \cite{Supplemental}. Notably, over half of the repumping lasers require less than 100~mW, a technically straightforward requirement.

In contrast to previous methods used to trap polyatomic molecules, which rely on strong electrostatic or magnetic fields \cite{zeppenfeld2012sisyphus, liu2017magnetic}, magneto-optical trapping (MOT) relies on rapid scattering of thousands of photons from lasers that are tuned into or out of resonance according to the molecules' position in a quadrupole magnetic field \cite{chu1998nobel}. This method leads to simultaneous trapping and cooling to near the Doppler limit. The optical cycle for CaOH is formed by bridging the $|N''=1, J''=1/2,-\rangle$ and $|N''=1, J''=3/2,-\rangle$ components of the electronic ground state manifold with a 52 MHz AOM (Fig. \ref{fig:Fig1}(b); hyperfine splittings in these states are $\leq2$ MHz and may be neglected). Both components are driven to the $|J'=1/2,+\rangle$ level of the $\widetilde{A}\,^2\Pi_{1/2}$ excited state, and their polarizations are chosen to preferentially scatter photons from the confining MOT laser beams (Fig. \ref{fig:Fig1}(b)) \cite{tarbutt2015magneto}. Magnetic dark states that occur in the ``type-II'' $|J''=1/2,3/2\rangle \rightarrow |J'=1/2\rangle$ transitions are remixed by rapidly switching laser polarizations between $\sigma^+/\sigma^-$ with a Pockels cell at 1.0 MHz. Trapping is maintained by modulating the current through the anti-Helmholtz MOT coils in phase with the polarization switching \cite{hummon20132d, anderegg2017radio, norrgard2016submillikelvin, baum20201d}. Each of the six MOT beams has a 1/$e^2$ Gaussian diameter of 10 mm, and the laser power in each beam is equally balanced between the $|N''=1,J''=3/2,-\rangle$ and $|N''=1,J''=1/2,-\rangle$ components. The axial rf magnetic field gradient has an RMS value of $\sim$8.5 G/cm.

The experimental sequence used to load the MOT is as follows. The slowing lasers are turned on 1 ms after ablation of the Ca target, and are held at constant power for a duration of 23 ms. During this time, the MOT beams are held at a power of 33 mW per beam with a detuning of $-2.0\Gamma$, where $\Gamma = 2\pi \times 6.4$ MHz is the natural linewidth of the cooling transition. When the slowing beam is turned off, the intensity in each MOT laser beam is ramped down to 3 mW over 5 ms, while the detuning is simultaneously ramped to $-0.9\Gamma$. All repumping lasers that are used for slowing (i.e. all except for the main cycling laser at 626 nm) are kept on for the full MOT sequence to address vibrational and rotational dark states inside the MOT. Thus, the same repumping lasers are used for repumping molecules during all three tasks of the experiment: slowing, MOT, and sub-Doppler cooling (see below).

\begin{figure}
    \centering
    \includegraphics[width=1\columnwidth]{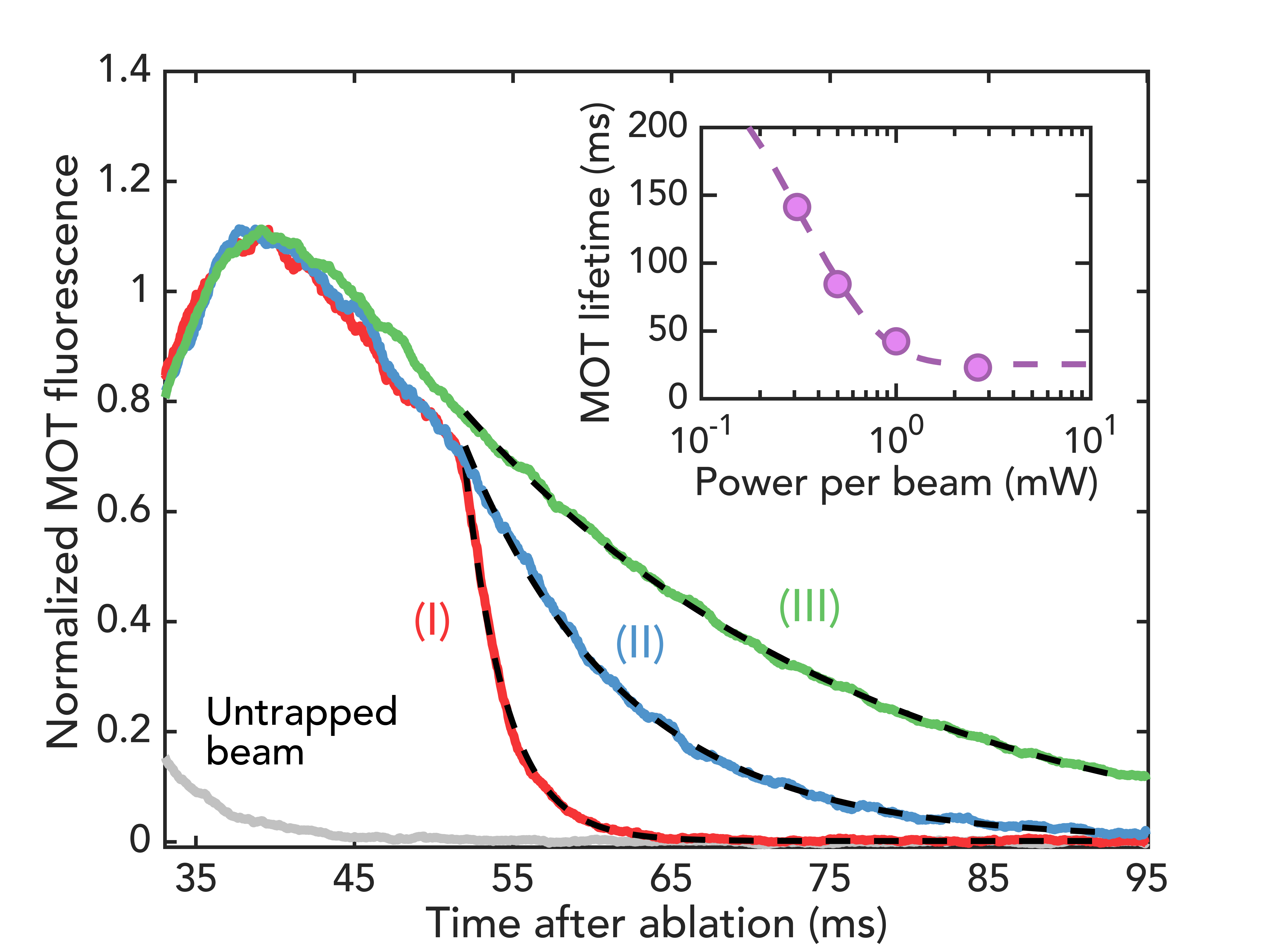}
    \caption{Time traces of fluorescence acquired by the PMT for an untrapped molecular beam (i.e. MOT off), and for MOT configurations with select repumpers switched off at 52 ms. The configurations achieve $1/e$ population loss after (I) 1200 (5 repumping lasers), (II) 4600 (8 repumpers), and (III) 12,000 scattered photons (11 repumpers). (I)-(III) are normalized to their mean signal from 35 to 52 ms. The untrapped beam is normalized to the average of (I)-(III) from 1 to 35 ms. Dotted lines are exponential fits, from which we extract MOT lifetimes of 2.60(3) ms, 10.1(2) ms, and 25.7(6) ms for (I)-(III). Errors are 1$\sigma$ fit errors. Inset: MOT lifetime versus power per MOT beam. The dotted line is a guide to the eye.
    }
    \label{fig:Fig2}
\end{figure}

To characterize the MOT, we collect fluorescence photons emitted at 555 nm resulting from spontaneous emission on the $\widetilde{B}\, ^2\Sigma^+(000) \rightarrow \widetilde{X}\,^2\Sigma^+(000)$ transition. Approximately one out of every 18 photons scattered during laser cooling occurs at this wavelength because decays to the first excited Ca--O stretching vibration are repumped via the $\widetilde{B}$ state (see Fig. \ref{fig:Fig1}(a)). This imaging scheme enables background-free detection due to the absence of cooling or repumping lasers at 555 nm \cite{Supplemental}. As shown in Fig. \ref{fig:Fig2}(a), the MOT manifests itself as a persistence of molecular fluorescence (collected on a photomultiplier tube (PMT)) at long times after the slowed molecular beam has fully passed through the trapping region. The signal is observed to decay exponentially due to molecules being lost from optical cycling into higher vibrational dark states. By selectively removing repumping lasers during trapping, the average number of photon scatters during cooling may be varied. We observe that the MOT lifetime depends linearly on the number of photons scattered, as expected \cite{Supplemental}. We extract from the lifetime measurements a scattering rate of $4.9(2)\times10^5$ s$^{-1}$ and determine from the magnitude of the collected fluorescence that $2.0(5)\times10^4$ CaOH molecules are loaded into the MOT \cite{Supplemental}. By decreasing the intensity of the MOT lasers beams after initial loading, the overall scattering rate can be reduced, extending the MOT lifetime beyond 100 ms (Fig. \ref{fig:Fig2} inset).

\begin{figure}
    \centering
    \includegraphics[width=1\columnwidth]{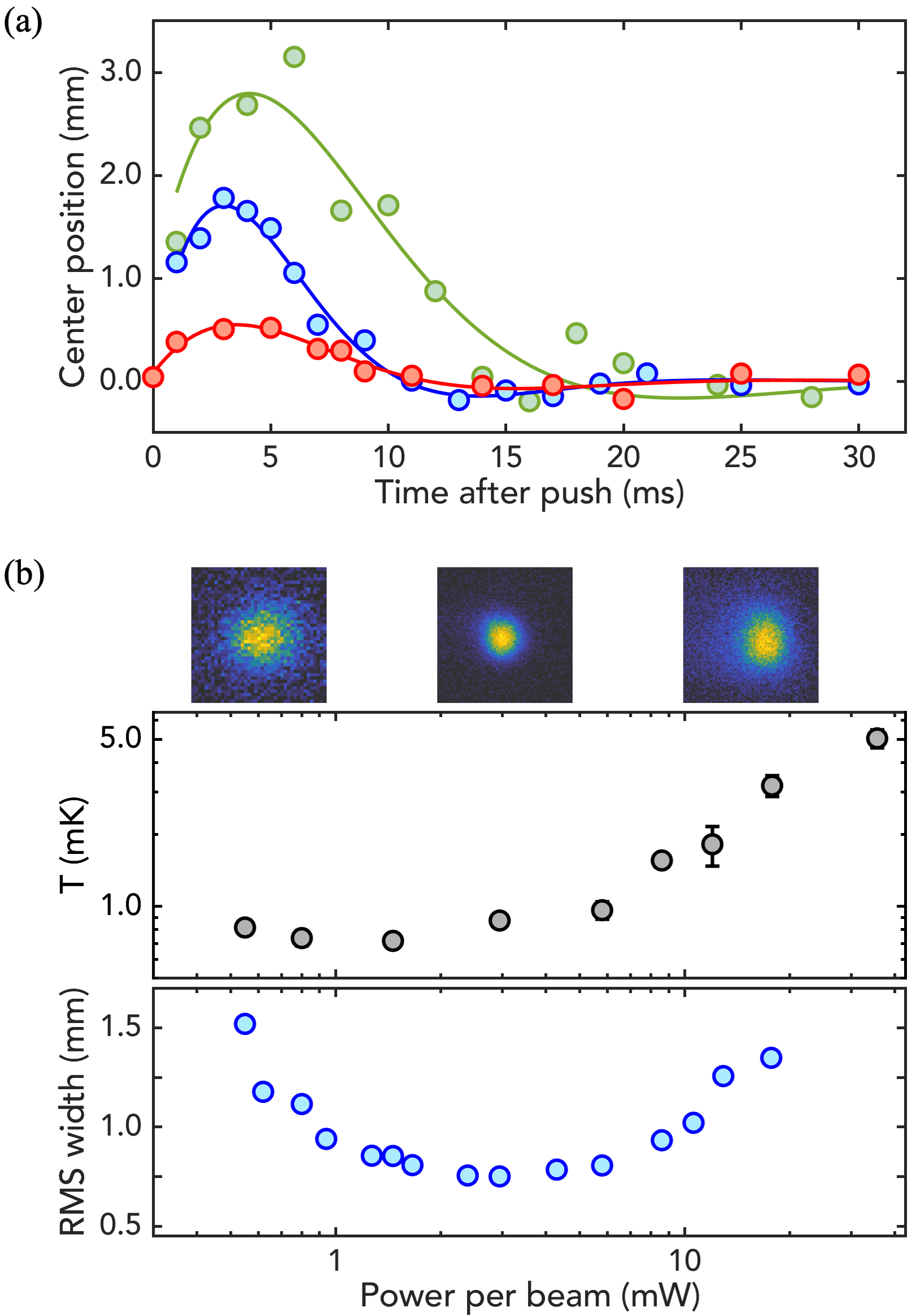}
    \caption{MOT dynamics and characteristics. (a) CaOH cloud position as a function of time after the pushing laser is applied, for MOT powers of 1.2 mW (green; top curve), 3 mW (blue; middle curve), and 11.5 mW (red; bottom curve). Fitted MOT parameters are $\beta=\{310(90),455(85),340(100)\}$ s$^{-1}$ and $\omega = 2\pi \times \{37(4),59(4),50(4)\}$ ms for the three powers, respectively. (b) MOT characteristics. Top: CaOH MOT images, from left to right, at powers of 0.5 mW, 3.0 mW, and 12.9 mW per MOT beam. The MOT is imaged for 10 ms following a 15 ms wait time to fully thermalize. The image size is 7 $\times$ 7 mm. Middle: Geometric mean temperature, $T=T_\text{radial}^{2/3}T_\text{axial}^{1/3}$, of the CaOH MOT versus laser power in each MOT beam. Bottom: Geometric mean, RMS width of the CaOH MOT as a function of laser power. Error bars are $1\sigma$ fit errors and are smaller than the plot markers for most points.}
    \label{fig:Fig3}
\end{figure}

Next, we measure the cooling of molecules and the damping of their motion in the MOT. MOT forces may be approximated at low velocity and displacement as those of a damped harmonic oscillator, $F/m = -\beta v - \omega^2 r$. Here $v$ and $r$ are the velocity and position in the trap, and $\beta$ and $\omega$ are the damping constant and trap frequency, respectively. To measure these properties, we observe the MOT after an intentional spatial displacement. After loading the MOT and ramping the MOT light to a power of 3 mW per beam, the slowing laser is turned on for 1 ms in order to displace the cloud of CaOH molecules from its equilibrium position. After a variable wait time, the molecules are imaged on an electron multiplying CCD (EMCCD) camera and the center position extracted (Fig. \ref{fig:Fig3}(a)). We measure a damping constant $\beta = 455(85)$ s$^{-1}$ and an oscillation frequency $\omega = 2\pi \times 59(4)$ Hz. The observed damping constant and frequency are comparable to those achieved in MOTs of diatomic molecules \cite{truppe2017molecules, anderegg2017radio}. Moreover, the characteristic damping time $\beta^{-1} \approx 2.9$ ms at this intensity is much shorter than the $1/e$ MOT lifetime of 26(1) ms (Fig. \ref{fig:Fig2}), indicating that significant cooling and compression of the MOT is possible. The damping is lower for MOT laser powers both above and below the optimal (see Fig. \ref{fig:Fig3}(a)), as expected \cite{williams2017characteristics}.

The temperature of the molecules in the MOT is measured by time-of-flight expansion. The MOT light is held at fixed power for $\sim$15 ms following the initial ramp, and then the MOT lasers and anti-Helmholtz coils are simultaneously turned off. After a variable time $\tau_\text{TOF}$, an image of the expanded molecular cloud is taken. The expanded cloud widths, $\sigma$, are determined from 2D Gaussian fits, and the radial and axial MOT temperatures are extracted from a linear fit of $\sigma^2$ vs. $\tau_\text{TOF}^2$ \cite{Supplemental}. The final MOT temperature depends strongly on the MOT power (Fig \ref{fig:Fig3}(b)) due to sub-Doppler heating effects \cite{devlin2016three}, as have been observed in other ``type-II'' MOTs \cite{norrgard2016submillikelvin,anderegg2017radio}. As the power is lowered, the temperature of the MOT reaches a minimum measured temperature of 720(40) $\mu$K. The diameter of the molecular cloud is smallest at a power of 3 mW per MOT beam (Fig. \ref{fig:Fig3}(c)), increasing at both lower and higher intensities. At 3 mW per MOT beam, a temperature of 870(50) $\mu$K, a peak number density of $3.0(8)\times 10^6$ cm$^{-3}$, and a peak phase-space density of $1.4(4)\times10^{-12}$ are achieved. These values are all comparable to those achieved in the first MOTs of diatomic molecules \cite{barry2014magneto,truppe2017molecules,anderegg2017radio,collopy20183d}. The MOT could be further improved by repumping additional, known, vibrational loss channels \cite{zhang2021accurate}.

\begin{figure}
    \centering
    \includegraphics[width=1\columnwidth]{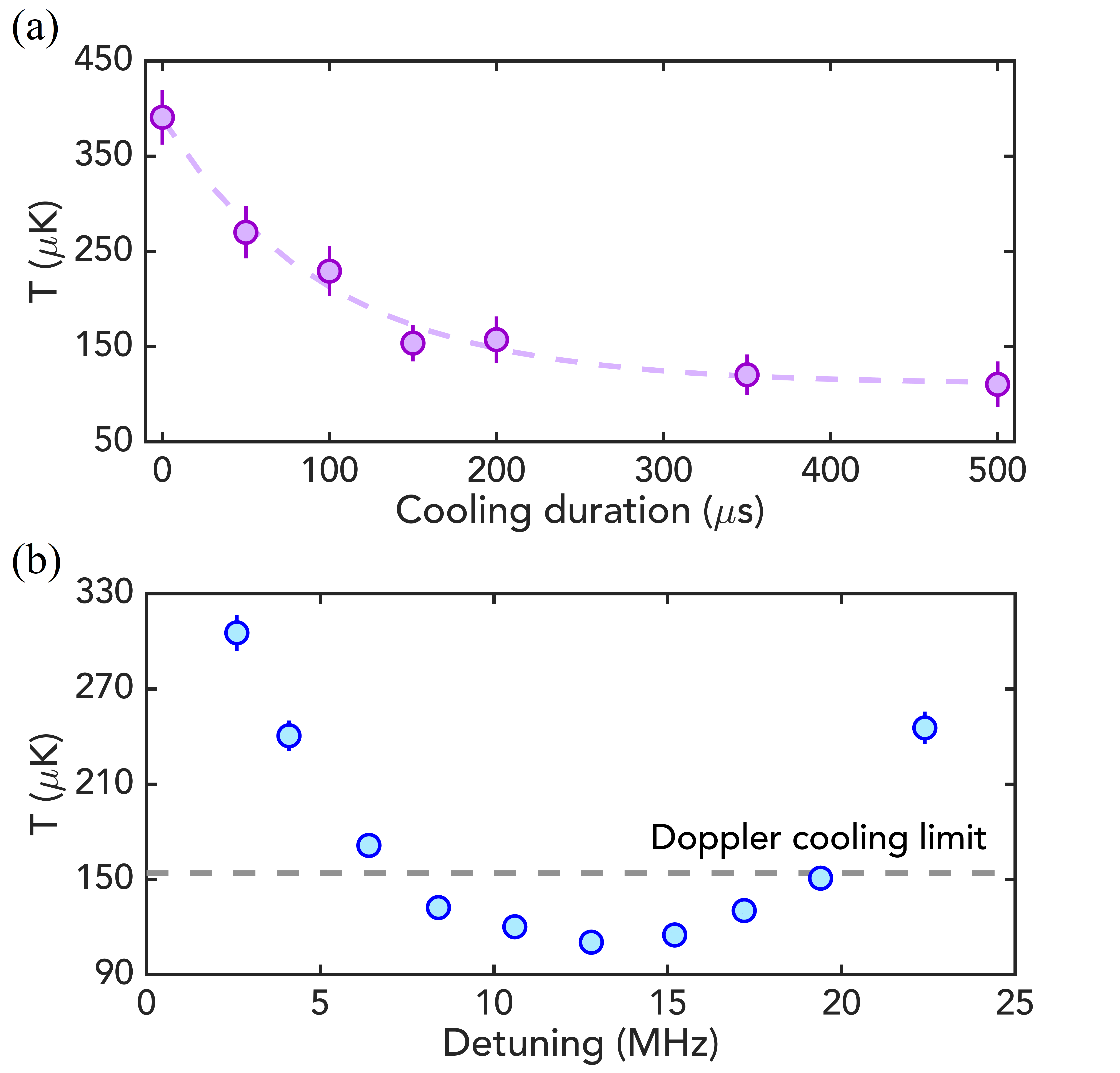}
    \caption{Sub-Doppler cooling. (a) Temperature as a function of cooling duration. The dotted line is an exponential fit, from which we extract a 1/$e$ cooling time of 136(11) $\mu$s. Error bars are 1$\sigma$ fit errors derived from 2D Gaussian fits of the expanded molecular cloud widths. (b) Temperature as a function of sub-Doppler detuning $\Delta_\text{SD}$. The Doppler cooling limit for CaOH is shown as a dotted line. Errors are calculated as for (a).}
    \label{fig:Fig4}
\end{figure}

We sub-Doppler cool the molecular cloud by applying a blue-detuned optical molasses. As demonstrated previously for diatomic molecules with the ``type-II'' $J\rightarrow J-1$ transitions used here \cite{truppe2017molecules, anderegg2018laser, lim2018laser}, the presence of dark states in the ground manifold leads to polarization gradient forces which result in strong sub-Doppler heating at red detuning and sub-Doppler cooling at blue detuning \cite{devlin2016three, devlin2018laser}. After loading and cooling the MOT, we perform sub-Doppler cooling by turning off the MOT coils, MOT lasers, and the polarization switching in a time period of 130 $\mu$s, while simultaneously jumping both frequency components of the MOT lasers to a variable detuning $\Delta_\text{SD}$. The cooling laser beams are then switched on for 1 ms at a power of 30 mW per beam. After this cooling pulse, the CaOH temperature is measured via time-of-flight expansion, as described above. The cooling occurs with a characteristic time of 136(11) $\mu$s (Fig. \ref{fig:Fig4}(a)) and a scattering rate of $7.7(4)\times10^5$ s$^{-1}$, corresponding to only 100 photon scatters for a $1/e$ decrease in temperature. We find an optimal detuning of $\Delta_\text{SD} = +13$ MHz (Fig. \ref{fig:Fig4}(b)). Since the lower frequency component of the sub-Doppler cooling light is red-detuned from the $|J' = 1/2\rangle \leftarrow |J''=1/2\rangle$ transition, sub-Doppler heating might be possible as the detuning increases. The minimum temperature acheived is 110(4) $\mu$K, which is comparable to the temperatures achieved in diatomic molecules using a similar sub-Doppler cooling mechanism \cite{truppe2017molecules, anderegg2017radio, caldwell2019deep}.

In summary, we have demonstrated magneto-optical trapping of a polyatomic molecule, CaOH, and performed sub-Doppler cooling to reach temperatures below the Doppler laser cooling limit. We expect that lower temperatures may be achieved by employing techniques that were used to sub-Doppler cool diatomic molecules to $\sim$10 $\mu$K \cite{cheuk2018lambda, caldwell2019deep, Langin2021}. At such temperatures, loading CaOH molecules into conservative optical traps will be possible \cite{anderegg2018laser, Langin2021}. Quantum control of individual CaOH molecules could then be achieved by loading into optical tweezer arrays, as demonstrated recently by work with calcium fluoride (CaF) molecules \cite{anderegg2019optical, Burchesky2021}. Such a platform would enable the ability to study and engineer interactions between individual, ultracold CaOH molecules, thus granting access to a wide range of potential experiments with polyatomic molecules in quantum simulation \cite{wall2013simulating, wall2015realizing, wall2014quantum} and quantum computation \cite{yu2019scalable}, in particular through the use of nearby opposite parity states in the vibrational bending mode \cite{kozyryev2017precision}. The demonstrated ability to rapidly scatter >$10^4$ photons in CaOH will greatly aid in achieving the high-fidelity state preparation and detection desired for these applications. Finally, this work is a proof of principle that managing rovibrational loss channels in polyatomic molecules for the purpose of laser cooling is both feasible and practical in species with optically accessible transitions and branching losses similar to those of CaOH. This is particularly notable as recent theoretical and experimental work indicates that these criteria are satisfied not only by linear, triatomic molecules, but also by several classes of molecules of even greater structural complexity. Several candidate species have already been identified, spanning nonlinear symmetric top molecules (e.g., CaOCH${_3}$, SrOCH${_3}$) \cite{kozyryev2016proposal}, asymmetric top molecules (e.g., CaSH, CaNH${_2}$) \cite{augenbraun2020molecular}, and complex organic molecules containing carbon rings (e.g., CaOC${_6}$H${_5}$, SrOC${_9}$H${_2}$F${_9}$) \cite{dickerson2021franck, dickerson2021optical}. 

We acknowledge Benjamin L. Augenbraun and Yicheng Bao for valuable discussions, and thank Svetlana Kotochigova and Jacek K\l os for providing theoretical insight into the CaOH chemical enhancement. This work was supported by the AFOSR and the NSF. NBV acknowledges support from the NDSEG fellowship, and LA from the HQI.

\bibliographystyle{apsrev}
\bibliography{CaOHMOT_references}

\end{document}